# 16-channel Photonic Solver for Optimization Problems on a Silicon Chip

Jiayi Ouyang[1], Shengping Liu[2], Ziyue Yang[1], Wei Wang[2], Xue Feng[1]*, Yongzhuo Li[1], and Yidong Huang[1]†

[1]Department of Electronic Engineering, Tsinghua University, Beijing 100084, China

[2]Chongqing United Microelectronics Center, Chongqing, China

Correspondence: * x-feng@tsinghua.edu.cn; † yidonghuang@tsinghua.edu.cn

**Abstract:** In this article, we proposed a programmable 16-channel photonic solver for quadratic unconstrained binary optimization (QUBO) problems. The solver is based on a hybrid optoelectronic scheme including a photonic chip and the corresponding electronic driving circuit. The photonic chip is fabricated on silicon on insulator (SOI) substrate and integrates high-speed electro-optic modulators, thermo-optic phase shifters and photodetectors to conduct the 16-dimensional optical vector-matrix multiplication (OVMM). Due to the parallel and low latency propagation of lightwave, the calculation of the QUBO cost function can be accelerated. Besides, the electronic processor is employed to run the heuristic algorithm to search the optimal solution. In the experiment, two 16-dimensional randomly generated QUBO problems are solved with high successful probabilities. To our knowledge, it is the largest scale of programmable and on-chip photonic solver ever reported. Moreover, the computing speed of the OVMM on photonic chip is ~2 TFLOP/s. It shows the potential of fast solving such optimization problems with integrated photonic systems.

**Introduction**

The rapid advances in artificial intelligence (AI) desire huge computation resources. Although the training of the large-scale neural networks is significantly improved with electronic hardware, *e.g.* graphics processing unit (GPU), the energy consumption is still heavy. Recently, various optical computing systems are proposed to demonstrate the optical neural networks (ONNs) due to the superiority of the high-speed and low-loss parallel propagation of lightwave. These platforms are based on spatial diffractive layers [1], on-chip optical matrix multiplications [2–4], on-chip diffractive optics [5,6], *etc*. Such optical computing systems exhibit high computation speed and energy efficiency in specific computation tasks, including the MNIST classification, vowel classification, alphabet classification, image classification, or content-generating. Besides the identification and classification, the non-deterministic polynomial hard (NP-hard) optimization problem is also important for AI technology. For the NP-hard optimization problem, it is required to find an input state to minimize the cost function. Such problems are ubiquitous but important issues in physics [7], finance [8], biology [9,10], *etc.*, including the subset sum problem [11,12], the Ising problem [13], and the travelling salesman problem (TSP) [14], *etc*. Solving NP-hard optimization problems can also be accelerated with optical computing systems as well [13], including those based on optical resonators [15–21], laser networks [22], and optical matrix multiplications [23–31], *etc*. However, considering the optical solver for optimization problems, the optical parameters have to be precisely configured to conduct a specific transmission function according to the given optimization problem, while the exact value of the transmission function is not so critical in neural networks. In addition, the ability of mapping arbitrary given problems demands high programmability and reconfigurability of the transmission function. Among the optical solvers, only a few of them are based on on-chip photonic systems. For instance, the subset sum problem can be solved with the photonic solver based on silicon photonic circuits consisting of split junctions [11,12], but the employed circuit is not reconfigurable and can only solve certain problems. In the photonic recurrent Ising sampler (PRIS), a programmable photonic circuit that can perform arbitrary optical vector-matrix multiplication (OVMM) is first employed to solve arbitrary Ising problem [26,27], which shows the feasibility of implementing the high-performance integrated photonic Ising machine. However, only four-dimensional Ising problems are experimentally solved with PRIS [27]. In the integrated photonic solver based on tunable delay lines, only the 5-node TSP problem is solved [14].

Although the previous photonic solvers have shown their potential to achieve high computing speed and energy efficiency, how to extend the dimensionality of the on-chip photonic solver still needs to be addressed. In this work, we proposed a 16-channel photonic solver for quadratic unconstrained binary optimization (QUBO) problems, which is based on a hybrid optoelectronic scheme including a photonic chip for the OVMM and the corresponding electronic driving circuit. With eigendecomposition of the weight matrix of the QUBO problem, the process of calculating the cost function could be accelerated with an on-chip OVMM. To conduct the required OVMM, the photonic chip integrates 16 optical amplitude modulators, a Mach-Zehnder interferometer (MZI) array including 64 multimode interference (MMI) couplers and 88 thermo-optic phase shifters as well as 16 balanced photodetectors (BPDs). The electronic driving circuit, which mainly comprises a field-programmable gate array (FPGA), is employed to control the photonic circuits, processes the optical signals, and conduct the necessary heuristic algorithm. In the experimental demonstration, two randomly generated QUBO problems with dimensionality of 16 are solved with high successful probability. To our knowledge, this is the largest scale of programmable and on-chip photonic solver for optimization problem. During the operation of the photonic solver, each single iteration takes ~265.1 ns, and the computing speed of the OVMM on photonic chip is ~2 TFLOP/s. These results indicate that the process of solving optimization problems can be accelerated with photonic chip.

## Results

**The architecture of the photonic solver.**

The proposed photonic QUBO solver employs a hybrid optoelectronic architecture with heuristic algorithms. The architecture of the solver is shown in Fig. 1, which consists of a laser operating at wavelength of 1550 nm, a photonic chip, and an electronic driving board. The photonic chip, which can perform the optical vector-matrix multiplication (OVMM), is fabricated on the silicon on insulator (SOI) substrate with the 130 nm process of silicon photonics in Chongqing United Microelectronics Center (CUMEC). The photonic chip consists of coupling gratings, beamsplitters, amplitude modulators, an MZI array, mixers and BPDs. Here, each electro-optic amplitude modulator is based on the MZI structure, which consists of two 1×2 MMI couplers and a phase shifter on each arm combined with meticulously designed travelling-wave electrodes and PN junction to achieve high modulation speed. Each mixer is actually a 2×2 MMI coupler. On the chip, the optical beam is first split into the signal and reference beams. The signal beam is further split into $N$ =16 beams as the input of the MZI array, while the detailed structure is presented in section "The principle of the OVMM". The driving board is composed of an FPGA (Xilinx Zynq UltraScale+ RFSoC XCZU29DR), digital-to-analog converters (DACs), operational amplifiers (OAs), and trans-impedance amplifiers (TIAs). The FPGA can control the bias voltages of the MZI array via a high-speed DAC to configure the 16×16 transformation matrix of the OVMM. Besides, the input vector is encoded and controlled by the FPGA via an DAC and amplitude modulators as denoted by the blue boxes in Fig. 1. After the MZI array, the optical signals would be detected and the current signals from the BPDs are converted to the voltage signals through the TIAs, which are received by the FPGA via an internal analog-to-digital converters (ADC). Furthermore, the FPGA executes the required procedure of the adopted heuristic algorithm. To obtain a stable OVMM, a cooling module is further employed on the photonic chip to maintain the temperature of the chip at 37°C during the operation.

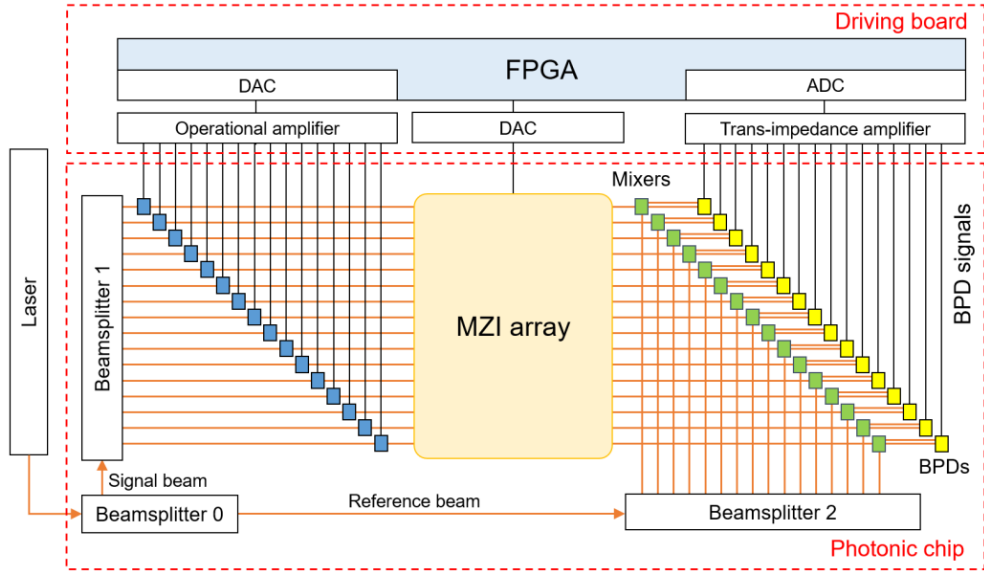

Fig. 1. The experimental setup of the photonic QUBO solver.

Our proposed photonic solver is employed to solve the NP-hard quadratic unconstrained binary optimization (QUBO) problem, where the $N$-length binary vector $\boldsymbol{s} \in \{0,1\}^N$ needs to be found to minimize the cost function $C(\boldsymbol{s})$ [32]:

$$C(\boldsymbol{s}) = -\frac{1}{2}\boldsymbol{s}^T \boldsymbol{K}\boldsymbol{s}. \qquad (1)$$

In Eq. (1), $\boldsymbol{K}$ is the real symmetric weight matrix and the superscript T denotes the transpose. At first, the weight matrix $\boldsymbol{K}$ can be decomposed by:

$$\boldsymbol{K} = \boldsymbol{Q}^T \boldsymbol{D} \boldsymbol{Q}, \qquad (2)$$

where $\boldsymbol{D} = \mathrm{diag}(\lambda_1, \lambda_2, \ldots, \lambda_N)$ is the diagonal eigenvalue matrix, and $\boldsymbol{Q}$ is the orthogonal eigenvector matrix. If all the eigenvalues of $\boldsymbol{K}$ are non-negative, $C$ can be written as

$$C = -\frac{1}{2}\boldsymbol{s}^T \boldsymbol{Q}^T \boldsymbol{D} \boldsymbol{Q}\boldsymbol{s} = -\frac{1}{2}\left(\sqrt{\boldsymbol{D}}\boldsymbol{Q}\boldsymbol{s}\right)^T \left(\sqrt{\boldsymbol{D}}\boldsymbol{Q}\boldsymbol{s}\right) = -\frac{1}{2}(\boldsymbol{A}\boldsymbol{s})^T (\boldsymbol{A}\boldsymbol{s}), \qquad (3)$$

where $A = \sqrt{D}Q$. In the OVMM, the input vector is set to $s$ and the transformation matrix is set to $A$. Denoting the output vector of the OVMM by $E_{\text{out}} = As$, Eq. (3) becomes

$$C = -\frac{1}{2}\sum_{i=1}^{N}\left(E_{\text{out}}^{(i)}\right)^2. \tag{4}$$

Eq. (4) indicates that the calculation of the cost function be accelerated by one step of OVMM. On the FPGA, the photonic solver executes a simulated-annealing-like algorithm *[30]* to search the near-optimal solutions of the QUBO problems. When solving a QUBO problem, the OVMM of the photonic chip is first configured to $A$ that corresponds to the weight matrix $K$. In each iteration, randomly selected $m \in [1,2,...,N]$ variables of the input vector $s$ is changed between 0 and 1, and the corresponding cost function is calculated with Eq. (10). Then the new state vector $s$ is accepted with the probability of $min[1, exp(\beta \Delta C)]$, where $\beta$ is the effective temperature and $\Delta C$ is the variation of $C$ relative to the last iteration. The higher $\beta$ is, the more $m$ is likely to take a smaller value. In the solving process, $\beta$ is gradually increased, and a near-optimal solution would be obtained when the algorithm terminates.

**The principle of the OVMM.**

As mentioned in the previous section, the photonic chip base on the pseudo-real-value architecture [33] can conduct the OVMM. The size of the photonic chip is 7.5×5 mm², as shown in the microscope image of Fig. 2. The 16-dimensional binary input vector $s$ is generated with the electro-optic amplitude modulators. Then the signal beams are guided to the MZI array. The structure of the FFT-mesh MZI array is shown in Fig. 3, which consists of 16 input ports, 4 layers of MZIs, and 16 output ports. In the MZI array, there are 88 thermo-optic phase shifters, which determine the transformation matrix. The cross connections between the MZIs are implemented with crossing waveguides. To enhance the spurious-free dynamic range of the MZI array, the "equal length, equal loss" design is employed, where the path and loss of light propagating from any input port to any output port are identical. Such design provides robustness against the environmental temperature fluctuation.

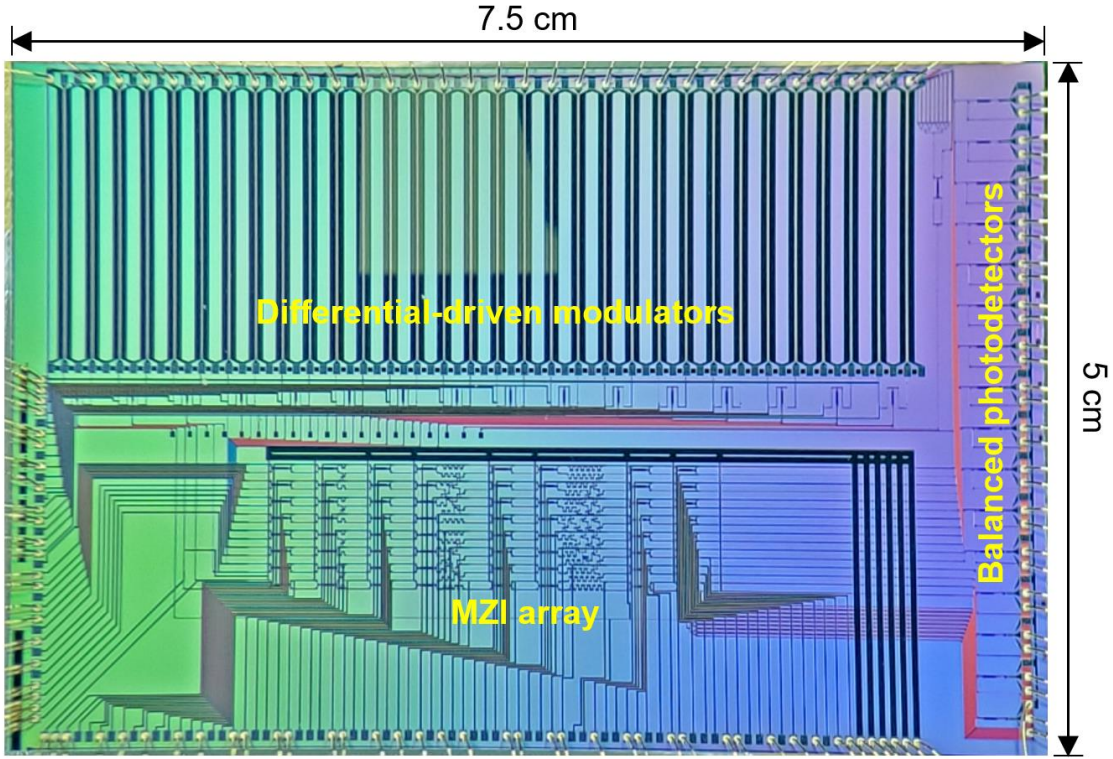

Fig. 2. The microscope image of the photonic chip to conduct the programmable OVMM.

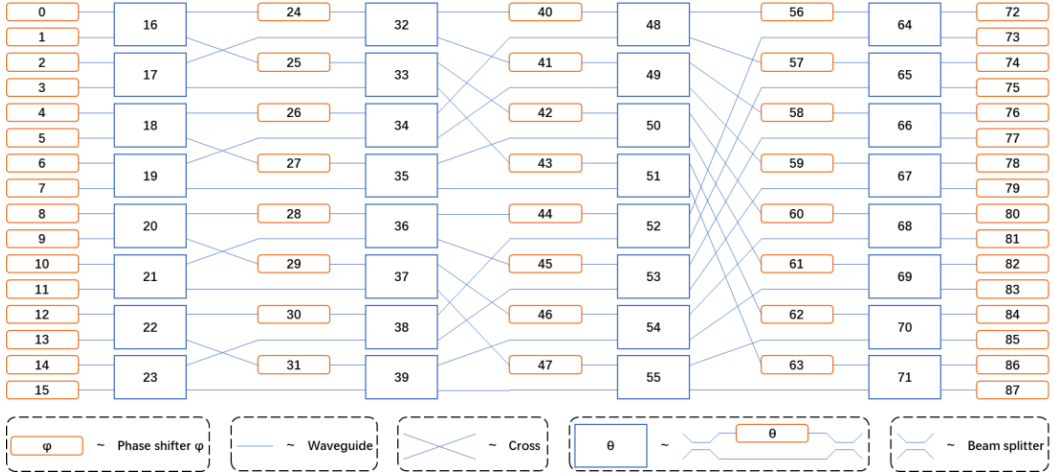

Fig. 3. The structure of the FFT-mesh MZI array. Each black box denotes a thermo-optic phase shifter. Each gray box represents an MZI consisting of two MMIs and two phase shifters.

Theoretically, the MZI array can perform a 16×16 unitary matrix transformation $U(V)$, which is a function of the bias voltages $V = [v_1, v_2 ... , v_{88}]$ on the thermo-optic phase shifters. The output complex amplitude vector of the MZI array is $E_{\text{out}} = Us$. As shown in Fig. 1, each output beam of the MZI array is input to a mixer along with 1 of 16 reference beams from beamsplitter 2. In Mixer $i$ ($i$ =1,2,…,16), which is also a beamsplitter shown in Fig. 3, we denote the complex amplitude of the beam from the MZI array by $E_{\text{out}}^{(i)} \exp(i\varphi_{\text{out}}^{(i)})$, and that of the reference beam is $E_{\text{ref}} \exp(i\varphi_{\text{ref}}^{(i)})$ (the amplitudes of the reference beams are the same). Therefore, the intensities of two output ports of mixer $i$ are

$$I_{\pm}^{(i)} = |E_{\text{ref}}|^2 + \left|E_{\text{out}}^{(i)}\right|^2 \pm |E_{\text{ref}}|\left|E_{\text{out}}^{(i)}\right| \cos(\varphi_{\text{out}}^{(i)} - \varphi_{\text{ref}}^{(i)}), \tag{5}$$

which would be then detected by BPD $i$, resulting in the differential current signal

$$I_{\text{BPD}}^{(i)} = I_{+}^{(i)} - I_{-}^{(i)} = 2|E_{\text{ref}}|\left|E_{\text{out}}^{(i)}\right| \cos(\varphi_{\text{out}}^{(i)} - \varphi_{\text{ref}}^{(i)}). \tag{6}$$

Since $\left|E_{\text{out}}^{(i)}\right| \cos(\varphi_{\text{out}}^{(i)}) = \text{Re}(E_{\text{out}}^{(i)})$ (Re(·) represents taking the real part), by setting all $\varphi_{\text{ref}}^{(i)}$ to 0,

$$I_{\text{BPD}}^{(i)} = 2|E_{\text{ref}}|\text{Re}(E_{\text{out}}^{(i)}) = 2|E_{\text{ref}}|\text{Re}(E_{\text{out}}^{(i)}). \tag{7}$$

In this work, the input vector is set to $s$ that consists of real elements $s_i \in \{0,1\}$. Therefore,

$$\text{Re}(E_{\text{out}}) = \text{Re}(Us) = \text{Re}(U)s, \tag{8}$$

and Eq. (7) becomes

$$I_{\text{BPD}} = 2|E_{\text{ref}}|\text{Re}(E_{\text{out}}) = 2|E_{\text{ref}}|\text{Re}(U)s. \tag{9}$$

Eq. (9) indicates that, if the BPD output signals $I_{\text{BPD}}$ is regarded as the output vector of the OVMM, the OVMM is indeed conducting a real matrix transformation $A \propto \text{Re}(U)$ on the input vector $s$. Such matrix $A$ is also a function of the bias voltages $A = A(V)$. To avoid confusions, in the following text, the output vector denotes the BPD signals, and the transformation matrix denotes the corresponding real transformation matrix $A$.

According to Eq. (3), any transformation matrix $A$ corresponds to a QUBO problem with weight matrix $K = A^T A$, which is positive semi-definite since $x^T K x = (Ax)^T A x \geq 0$ for any $x \in \mathbb{R}^N$. Thus, according to Eq. (3) and (4), the experimental cost function $C_{\text{exp}}$ can be calculated from the output signal

$$C_{\text{exp}} = -\frac{1}{2}\sum_{i=1}^{N}\left(I_{\text{BPD}}^{(i)}\right)^2. \tag{10}$$

**Experiment.** In the experiment, we generated two configurations of the bias voltages $V_1$ and $V_2$ (the 32 voltages on the MZIs in Fig. 3 are randomly generated while the others are 0), and then employed them on the photonic chip respectively. Thus, two transformation matrices $A_1$ and $A_2$ can be measured respectively, which are shown in Fig. 4. The transformation matrices $A_1$ and $A_2$ corresponds to the QUBO problems with weight matrices $K_1$ and $K_2$ respectively. The two problems are denoted by Q1 and Q2 for simplicity, and their weight graphs are shown in Fig. 5 respectively, where the black dots denote the variables and the color of the connection denotes the weight.

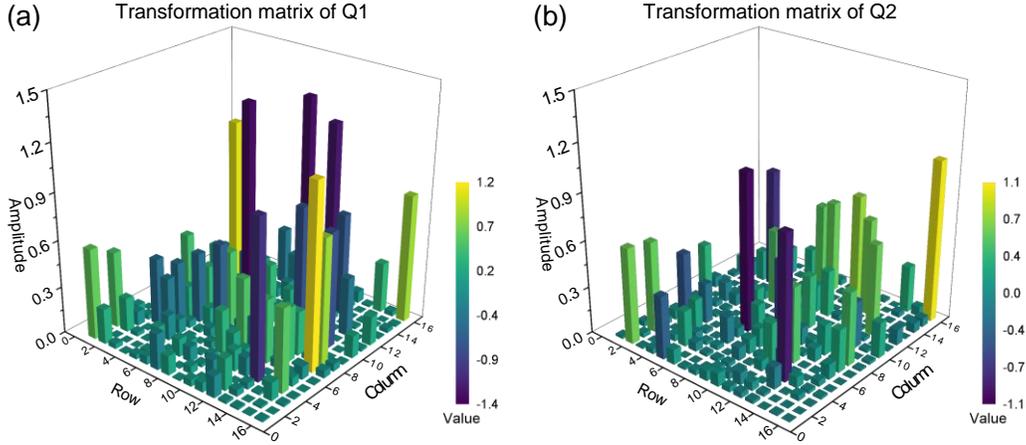

Fig. 4. The randomly generated transformation matrices. (a) $A_1$ of Q1. (b) $A_2$ of Q2.

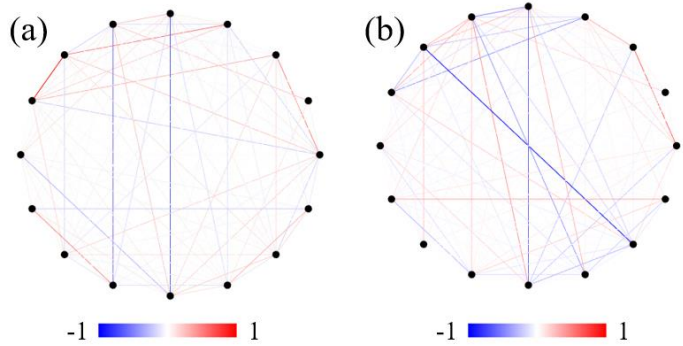

Fig. 5. The weight graphs of the demonstrated problems. (a) The weight graph of Q1. (b) The weight graph of Q2.

Each problem is solved for 100 times, and the accepted state $s$ in all iterations are recorded. To evaluate the solving performance, the cost function $C$ of the accepted states are calculated with Eq. (1), and the evolution curves of $C$ of Q1 and Q2 are plot in Fig. 6(a) and 6(b) respectively. It can be seen that $C$ fluctuates above $C_{\min}$ denoted by the black dashed line ($C_{\min}$ is the lowest $C$ obtained with numerical simulations) within a small range after 400 iterations. To quantify the solving performance, we consider one iteration is successful if the theoretical $C$ of the accepted state vector $s$ is lower than $\eta C_{\min}$, where $\eta \in (0,1]$ is the tolerance coefficient. The high successful probability of the last iteration when $\eta$ is close to 1 can indicate the high searching accuracy. Thus, the successful probability of each iteration under $\eta = 0.96 \sim 0.99$ is calculated, and the results of Q1 and Q2 are shown in Fig. 6(c) and 6(d) respectively. Since the initial state of each run is randomly generated, the successful probability of the first iteration is 0. Then the successful probability grows rapidly from iteration 1 to 400, and finally fluctuates around a stable value. It can be seen from Fig. 6(c) and 6(d) that, the final successful probability is close to 1 when $\eta \leq 0.98$, which indicates that our photonic solver is capable of solving the QUBO problems.

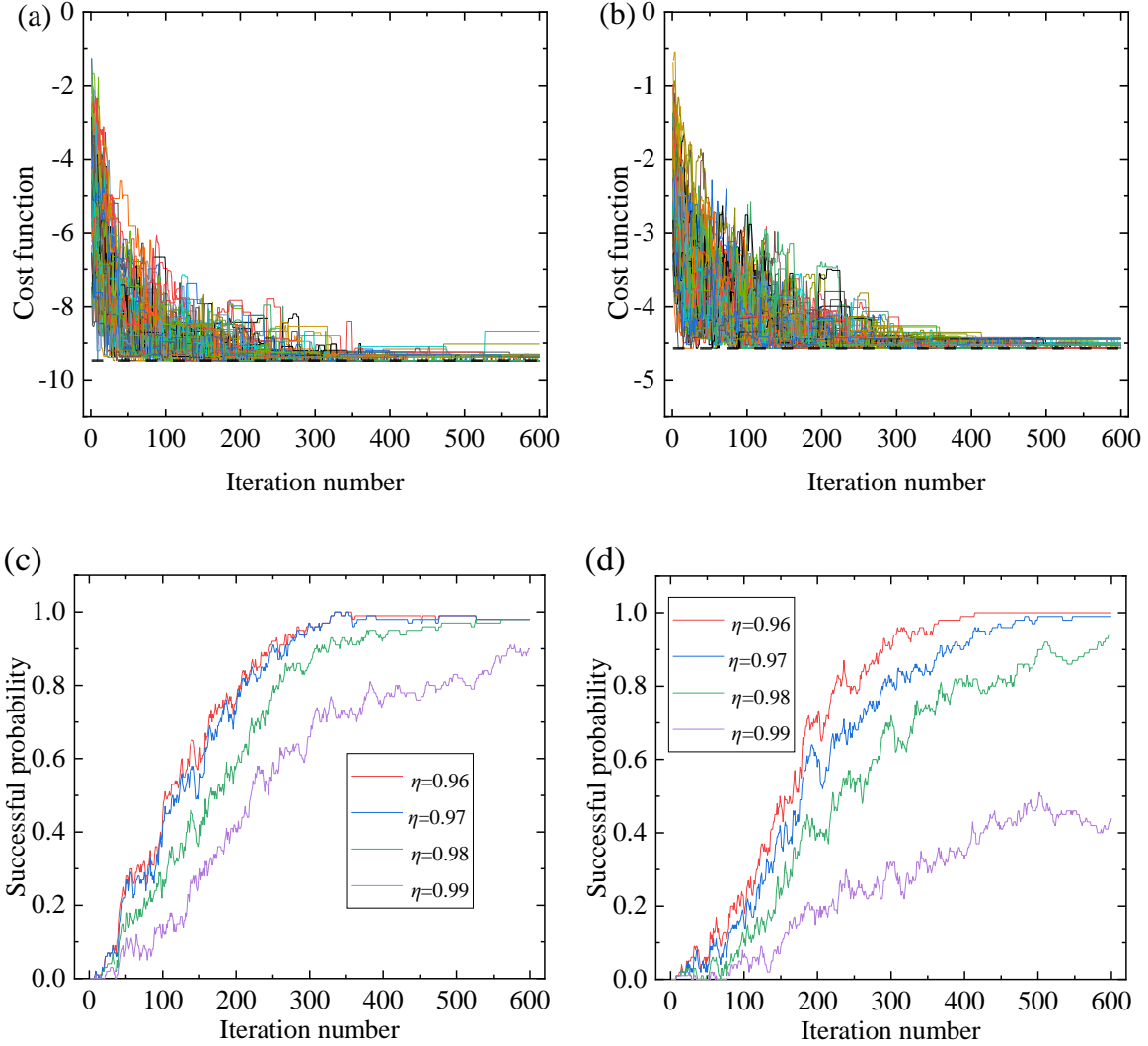

Fig. 6. The experimental results. (a) and (b) are the normalized experimental Hamiltonian evolution curves of Q1 and Q2 respectively. (c) and (d) are the successful probability curves of Q1 and Q2 respectively.

However, it can be observed from Fig. 6(a) and 6(b) that the final successful probability would decay when $\eta > 0.98$. We believe that such decay is mainly due to the fluctuation of the experimental cost function, which may result from the detection noise of the photodetector, the deviation of the OVMM, and the fluctuation of the laser power. We will discuss them in the next section.

**The accuracy of the OVMM.** We first discuss the stability of the OVMM, which is directly associated with the accuracy of the experimental cost function. Due to the high sensitivity to the temperature of the photonic chip, even a small temperature variation would result in a completely different optical matrix transformation. Therefore, a cooling module is employed to stabilize the working temperature of the photonic chip at 37 °C. In the experiment, the temperature fluctuation is less than 0.005°C. To quantify the stability of the OVMM, the fidelity $F$ of each output vector is calculated with

$$F = \frac{\left|\sum_i I_{\text{BPD}}^{(i)} I_{\text{T}}^{(i)}\right|}{\sqrt{\sum_i \left[I_{\text{BPD}}^{(i)}\right]^2 \sum_i \left[I_{\text{T}}^{(i)}\right]^2}}, F \in [0,1], \tag{11}$$

where $\boldsymbol{I}_{\text{T}} = \left[I_{\text{T}}^{(i)}\right] = \boldsymbol{As}$ is the theoretical output vector. The fidelity indicates the parallelism of the vectors $\boldsymbol{I}_{\text{BPD}}$ and $\boldsymbol{I}_{\text{T}}$. The fidelity close to 1 can indicate the high stability of the OVMM. The average fidelity and the corresponding standard deviation of each run for Q1 and Q2 is shown in Fig. 7(a) and 7(b), respectively. It can be seen that all the average fidelities are higher than 0.99 with small distribution, and the total average fidelities of Q1 and Q2 are 0.9953±0.0017 and 0.9934±0.0025 respectively. Such result indicates that the OVMM of the photonic chip is quite stable during the solving process.

Besides the fidelity, the noise of the BPD and the fluctuation of the laser power can also degrade the accuracy of the OVMM. If the total noise level of the cost function is larger than the variation of the cost function, the photonic solver would give the incorrect solution. Therefore, the scale factor of the experimental cost function is used to evaluate the total noise level. The scale factor $P$ in each iteration is defined as

$$P = \frac{C_{\text{exp}}}{C_{\text{theo}}} = \frac{|I_{\text{BPD}}|^2}{|I_{\text{T}}|^2}, \quad (12)$$

where $C_{\text{theo}} = -|I_{\text{T}}|^2/2$ is the theoretical cost function. In ideal conditions, $P$ should be a constant. In the experiment, the fluctuation of $P$ mainly results from the laser power fluctuation and the detection noise. When $C_{\text{theo}}$ are equal, from Eq. (12) we have

$$\text{SNR} = \frac{C_{\text{exp}}}{\Delta C_{\text{exp}}} = \frac{P}{\Delta P}, \quad (13)$$

where SNR denotes the signal-to-noise ratio of the experimental cost function, $\Delta C_{\text{exp}}$ is the noise level of $C_{\text{exp}}$, and $\Delta P$ is the standard deviation of the scale factor $P$. For convenience, the average scale factor and the corresponding standard deviation of each run is calculated with $P$ in all iterations, and the results of Q1 and Q2 are shown in Fig. 7(c) and 7(d) respectively. It can be seen that, in each run the scale factor $P$ only distributes in a small range. The average SNRs of 100 runs of Q1 and Q2 are 26.6 dB and 28.2 dB respectively, which indicates that the noise level of the experimental cost function is small. The inverse of the SNR can also be regarded as the "resolution" of the cost function $R = 1/\text{SNR}$, hence the corresponding $R$ of Q1 and Q2 are 4.67% and 3.88% respectively. Such parameter indicates that the photonic solver cannot distinguish two states if $R > |\Delta C/C_{\text{min}}|$, where $\Delta C$ is the difference of their cost functions and $C_{\text{min}}$ is the minimum cost function.

To further investigate how much the noise level would affect the searching process near the ground state, the theoretical relative variation of the cost function, $C_{\text{r}} = \Delta C/C_{\text{min}}$, between the sampled state and the previous state of all samplings are calculated in iteration 400-600. Obviously, if $R > C_{\text{r}} > 0$ in a single iteration, though the theoretical cost function increases, which means the sampled state would not be accepted, the experimentally measured cost function might decrease due to the noise. Thus, a wrong acceptance of the sampled state would occur. Among these samplings, we found that the proportions of the samples that satisfy $R > C_{\text{r}} > 0$ are only 2.37% and 4.30% for model 1 and 2, respectively. It indicates that our photonic solver can distinguish two states in most iterations, hence high successful probabilities can be obtained even when the tolerance coefficient is close to 1. The above analysis indicates that our OVMM is quite stable and accurate during the solving process.

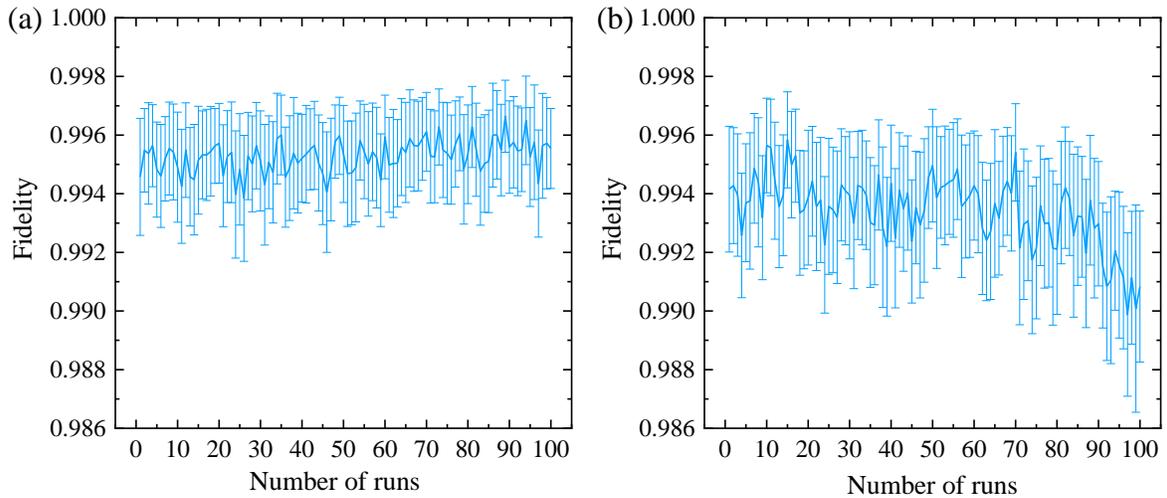

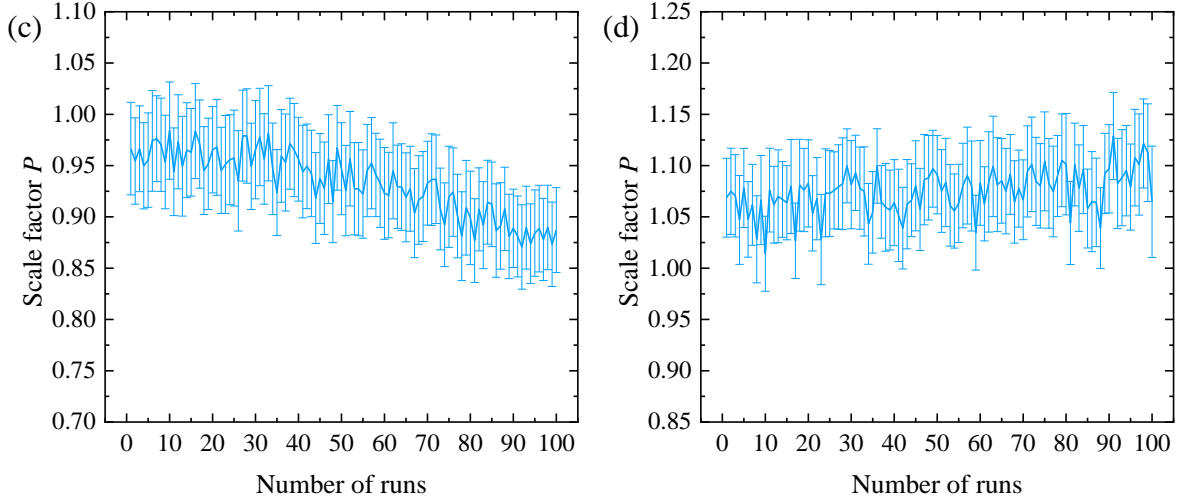

Fig. 7. The stability analysis of the OVMM in the experiment of Q1 and Q2. (a) and (b) show the average fidelities of each runs of Q1 and Q2 respectively. (c) and (d) show the average amplitude stability factor $P$ of each runs of Q1 and Q2 respectively.

**Computation speed.** As mentioned in the previous section, the calculation of the photonic solver includes the pretreatment and the iterative sampling of the state vector. In the pretreatment, the eigendecomposition of the weight matrix of the given problem is calculated to obtain the necessary transformation matrix of the OVMM. In the experimental demonstration, the transformation matrices of the OVMM are randomly generated, hence we only discuss the time consumption in the iterative sampling process.

The main sequence diagram of each sampling iteration is shown in Fig. 8. The clock rate of the photonic chip and the FPGA is 245.76 MHz, corresponding to the clock cycle of $t_0$ =4.069 ns. At $t = 0$, the FPGA sends the TX (transmit) signal containing the voltages on the amplitude modulators to the DAC. We observed that the RX (receive) signal appears at $t = 40t_0$. Such latency is composed of five parts in sequence: the DAC latency, the modulator response time, the propagation time of lightwave in the OVMM, the photodetector response time, and the ADC latency. According to the frequency responses of the modulator and the photodetector shown in Fig. 9(a) and 9(b) respectively, the -3 dB bandwidth $f_B$ of the amplitude modulator is ~28.0 GHz and that of the photodetector at -2 V bias voltage is ~41.3 GHz. Therefore, the response times $\tau_r$ of the modulator and the photodetector can be estimated to be 12.5 ps and 8.5 ps respectively according to $\tau_r \approx 0.35/f_B$ [34]. The length from an amplitude modulator to a photodetector is ~9.3 mm, and the refractive index of silicon under the working temperature of 37 °C is ~3.48 [35], hence the propagation time of light is ~107.9 ps. Thus, the computing of the OVMM takes $\tau_{OVMM} \approx$ 128.9 ps in each iteration, and the total latency of the DAC and ADC can be estimated as $\tau_{DAC/ADC} \approx$ 162.6 ns, which indicates that $\tau_{OVMM}$ is much lower than $\tau_{DAC/ADC}$. After some smoothing and averaging operations, the processed RX signal appears at $t = 45t_0$, and the sampled signal rises at $t = 47t_0$ and falls at $t = 51t_0$. According to the experimental results, each iteration takes $\tau_{iter} \approx$ 265.1 ns. Thus, the time consumption of the heuristic algorithm on the FPGA takes $\tau_{FPGA} = \tau_{iter} - 51t_0 \approx$ ~57.6 ns.

In each iteration, calculating the cost function requires the multiplication between an $N \times N$ matrix and an $N \times 1$ vector, which includes $N^2$ floating-point operations (FLOPs). After considering the DAC/ADC latency, the speed of the whole multiplication on chip is ~1.57 GFLOP/s. Actually, the speed of required vector-matrix multiplication is mainly limited by the the DAC and ADC. As mentioned above, the computing time of the OVMM ($\tau_{OVMM} \approx$ 128.9 ps) is much lower than the total latency of the DAC and ADC ($\tau_{DAC/ADC} \approx$ 162.6 ns). The computation speed of OVMM on photonic chip can be estimated as 2.00 TFLOP/s. Correspondingly, the area efficiency of the photonic chip is ~53.3 GMAC/mm$^2$ with the chip area of 37.5 mm$^2$ (the length/width is 7.5/5 mm). The area efficiency of the photonic chip can be further improved with the more compact chip size via advanced chip fabrications. From the analysis above, it can be seen that the bottleneck of the computation speed of our photonic solver is the DAC/ADC latency. The low-latency DAC/ADC can be employed to further shorten the iteration time and speed up the computation. For instance, the latencies of the DAC DAC5670 (Texas Instruments) and ADC ADS52790 (Texas Instruments) are ~3.5 ns and ~3.4 ns, which would reduce the time consumption from the DAC to the ADC to ~7.1 ns. The FPGA with higher main frequency would also be helpful to reduce the iteration time.

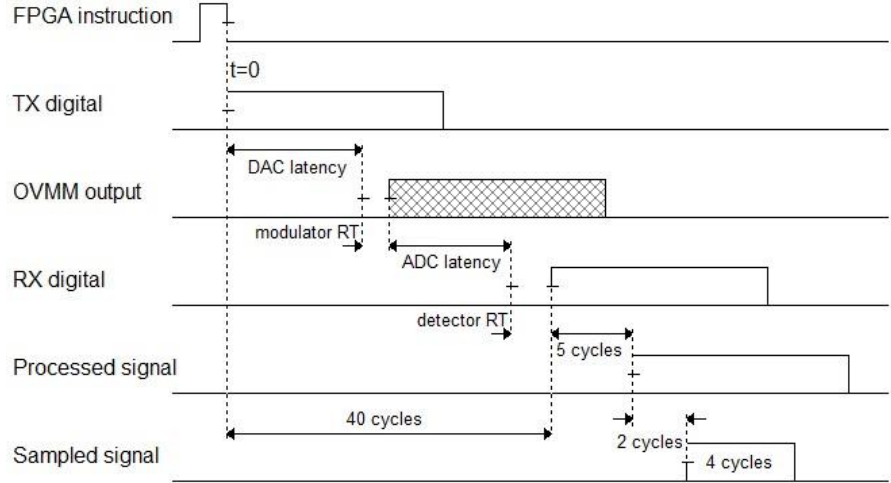

Fig. 8. Main sequence diagram of each sampling iteration. RT: response time.

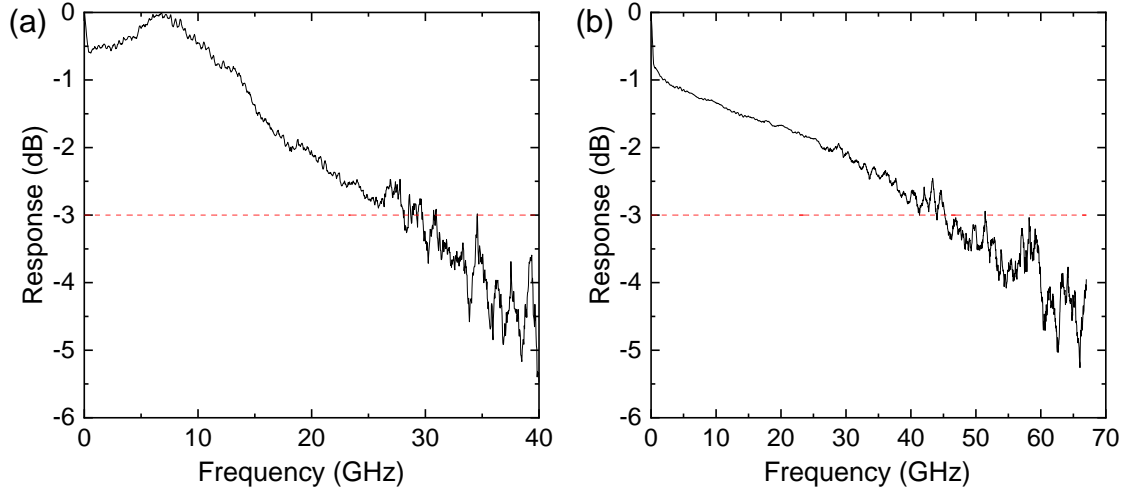

Fig. 9 The frequency responses of the amplitude modulators and the photodetectors. (a) The result of the amplitude modulators. (b) The result of the photodetectors.

**Conclusion**

In this article, we have proposed and demonstrated a 16-channel photonic solver based on an integrated photonic chip for QUBO problems. The amplitude modulators, the MZI array for the optical matrix transformation, and the photodetectors are all integrated on the same photonic chip to achieve low optical latency. Calculating the cost function of the QUBO problem can be accelerated with the photonic chip while heuristic algorithms are employed on the FPGA to search the optimal solution. Two randomly generated 16-dimensional QUBO problems have been successfully solved with the photonic solver, and the final successful probabilities of both problems are larger than 0.94 with the tolerance coefficient of 0.98. Such results indicate the high stability and low noise level of the photonic solver. According to the experimental results, each iteration takes ~265.1 ns, which is mainly contributed by the DAC/ADC latency (~162.6 ns). The computing speed and the area efficiency of the photonic chip are 2.00 TFLOP/s and 53.3 GMAC/mm$^2$, respectively. Therefore, the proposed photonic solver shows the potential of the integrated photonic system to accelerate the solving process of computationally complex problems. It also should be mentioned that the computing speed of the photonic solver can be further improved with low-latency DAC/ADC.

[28] D. Pierangeli, M. Rafayelyan, C. Conti, and S. Gigan, *Scalable Spin-Glass Optical Simulator*, Phys. Rev. Appl. **15**, 034087 (2021).
[29] H. Yamashita, K. Okubo, S. Shimomura, Y. Ogura, J. Tanida, and H. Suzuki, *Low-Rank Combinatorial Optimization and Statistical Learning by Spatial Photonic Ising Machine*, Phys. Rev. Lett. **131**, 063801 (2023).
[30] J. Ouyang, Y. Liao, Z. Ma, D. Kong, X. Feng, X. Zhang, K. Cui, F. Liu, W. Zhang, and Y. Huang, *An On-Demand Photonic Ising Machine with Simplified Hamiltonian Calculation by Phase Encoding and Intensity Detection*, (n.d.).
[31] J. Ouyang, Y. Liao, X. Feng, Y. Li, K. Cui, F. Liu, H. Sun, W. Zhang, and Y. Huang, *A Programmable and Reconfigurable Photonic Simulator for Classical XY Models*, arXiv:2401.08055.
[32] G. Kochenberger, J.-K. Hao, F. Glover, M. Lewis, Z. Lü, H. Wang, and Y. Wang, *The Unconstrained Binary Quadratic Programming Problem: A Survey*, J Comb Optim **28**, 58 (2014).
[33] Y. Tian, Y. Zhao, S. Liu, Q. Li, W. Wang, J. Feng, and J. Guo, *Scalable and Compact Photonic Neural Chip with Low Learning-Capability-Loss*, Nanophotonics **11**, 329 (2022).
[34] Z.-Y. Li, D.-X. Xu, W. R. McKinnon, S. Janz, J. H. Schmid, P. Cheben, and J.-Z. Yu, *Silicon Waveguide Modulator Based on Carrier Depletion in Periodically Interleaved PN Junctions*, Opt. Express, OE **17**, 15947 (2009).
[35] H. H. Li, *Refractive Index of Silicon and Germanium and Its Wavelength and Temperature Derivatives*, Journal of Physical and Chemical Reference Data **9**, 561 (1980).
[36] M. Reck, A. Zeilinger, H. J. Bernstein, and P. Bertani, *Experimental Realization of Any Discrete Unitary Operator*, Physical Review Letters **73**, 58 (1994).